\documentclass[pra,twocolumn]{revtex4}

\usepackage{amsmath}
\usepackage{color}
\usepackage{dcolumn}
\usepackage[cp1251]{inputenc}
\usepackage[english]{babel}
\usepackage{epsfig}
\usepackage{array}

\begin{document}
 \title{Biphoton Double-Bell States and Ququarts with ``Invisible$"$ Variables}

\author{M.V. Fedorov$^{1\,*}$, P.A. Volkov$^{1}$, J.M. Mikhailova$^{1,\,2}$,}
\affiliation{$^1$A.M.Prokhorov General Physics Institute, Russian Academy of Science, Moscow, Russia\\
$^2$Max-Planck-Institut f\"ur Quantenoptik, Garching, Germany\\
$^*$e-mail: fedorov@gmail.com}

\begin{abstract}

We analyze features of mixed biphoton polarization states which arise from pure states of polarization-frequency biphoton ququarts after averaging over frequencies of photons. For mixed states we find their concurrence $C$, Schmidt parameter $K$ and degree of polarization $P$, as well as the von Neumann mutual information $I$. In some simple cases we find also the relative entropy $S_{rel}$ and the degree of classical correlations $C_{cl}$. We show that in mixed states the Schmidt parameter does not characterize anymore the degree of entanglement, as it does in pure states. Nevertheless, the Schmidt parameter remains useful even in the case of mixed states because it remains directly related to the degree of polarization. We compare results occurring in the cases of full pure polarization-frequency states of ququarts, mixed states (averaged over frequencies) and states with separated high- and low-frequency parts. Differences between these results can be seen in experiments with and without a dichroic beam-splitter, as well as with and without frequency filters in front of detector.

\end{abstract}

\maketitle

\section{Introduction}

As known, a pure state of two particles is entangled if its bipartite wave function cannot be factorized, i.e., presented in the form of  a product of two single-particle functions,
\begin{equation}
 \label{SCHR}
 \Psi(x_1,x_2)\neq \varphi(x_1)\times \chi(x_2),
\end{equation}
where $x_1$ and $x_2$ are variables or sets of variables of two particles. A natural extension of this definition is given by the Schmidt decomposition, or Schmidt theorem \cite{Grobe,Ekert}, according to which any entangled, unfactorable, bipartite wave function can be presented as a sum of factorized terms
\begin{equation}
 \label{SCHMIDT-MODES}
 \Psi(x_1,x_2)=\sum_n\sqrt{\lambda_n}\;\varphi_n(x_1)\times \chi_n(x_2).
\end{equation}
The orthogonal and normalized functions $\varphi_n(x_1)$ and $\chi_n(x_2)$ are the Schmidt modes defined as eigenfunctions of the reduced density matrices $\rho_r^{(1)}(x_1,x_1^\prime)$ and $\rho_r^{(2)}(x_2,x_2^\prime)$, which are defined in their turn as partial traces of the full density matrix $\rho(x_1,x_2;x_1^\prime,x_2^\prime)=\Psi(x_1,x_2)\Psi^*(x_1^\prime,x_2^\prime)$ over $x_2$ (for $\rho_r^{(1)}$) or over $x_2$ (for $\rho_r^{(2)}$); $\lambda_n$ denotes coinciding eigenvalues of  both reduced density matrices $\rho_r^{(1)}$ and $\rho_r^{(2)}$, $n=1,\,2,\,...$  The amount of efficiently populated Schmidt modes in the Schmidt decomposition (\ref{SCHMIDT-MODES}) characterizes the degree of entanglement and can be evaluated by the Schmidt parameter
\begin{equation}
 \label{K}
 K=\frac{1}{Tr_{x_1}\rho_r^{(1)\,2}}=\frac{1}{Tr_{x_2}\rho_r^{(2)\,2}}
 =\frac{1}{\sum_n\lambda_n^2}.
\end{equation}
For shortening formulations, let us refer the characterization of entanglement by the Schmidt decomposition (\ref{SCHMIDT-MODES}) and the Schmidt parameter $K$ (\ref{K}) as the Schmidt entanglement. Note also that the Schmidt decomposition and the definition of the Schmidt parameter (\ref{K}) are valid equally for bipartite systems with either discrete or continuous variables $x_{1,2}$.

It is rather important to analyze the relation of these definitions with those arising in the occupation-number representation of bipartite states. In the occupation-number representation, states are characterized by modes $i=1,\,2,\,...N$, and numbers of particles in these modes, $n_i$. In the case of bipartite states $\sum_in_i=2$. Each specific distribution of particle occupation numbers in modes can be referred to as a configuration and denoted as $\{n_i\}$. The bipartite density matrix in the representation of occupation numbers is defined by its matrix elements  $\langle \{n_i\}|\hat{\rho}|\{n_i^\prime\}\rangle$. On the other hand, each particle has its ``coordinate$"$, or sets of coordinates, $x_1$ and $x_2$, correspondingly, for particles 1 and 2. Coordinate matrix elements of the density matrix $\hat{\rho}$ between the coordinate eigenstates detrmine the coordinate density matrix $\rho(x_1,x_2;x_1^\prime,x_2^\prime)=\langle x_1,x_2|\hat{\rho}|x_1^\prime,x_2^\prime\rangle$. The sets of states $|x_1,x_2\rangle$ and $|\{n_i\}\rangle$  are equally complete, and there is the following relation between the density matrix in the coordinate and occupation-number representations
\begin{gather}
 \nonumber
 \rho(x_1,x_2;x_1^\prime,x_2^\prime)
  = \sum_{\{n_i\},\{n_i^\prime\}} \langle x_1,x_2|\{n_i\}\rangle\\
 \label{dens-matr-multimode}
 \times\langle \{n_i\}|\hat{\rho}|\{n_i^\prime\}\rangle\;
 \langle\{n_i^\prime\}|x_1^\prime,x_2^\prime\rangle.
\end{gather}
The transformation coefficients have sense of the bipartite coordinate wave functions for given configurations
\begin{gather}
 \nonumber
 \langle x_1,x_2|\{n_i\}\rangle=\Psi_{\{n_i\}}(x_1,x_2),\\
 \label{conf-wf}
 \langle\{n_i^\prime\}|x_1^\prime,x_2^\prime\rangle=
 \Psi_{\{n_i^\prime\}}^*(x_1^\prime,x_2^\prime).
\end{gather}
In the case of distinguishable particles, the given-configuration wave functions (\ref{conf-wf}) are always factorized, and the Schmidt entanglement can arise only if the sum over configurations in Eq. (\ref{dens-matr-multimode}) contains more than one term. This case can be referred to as corresponding to the configuration entanglement. On the other hand, in the case of indistinguishable particles, the given-configuration wave functions (\ref{conf-wf}) must be either symmetric or antisymmetric with respect to the variable transpositions (correspondingly, in the cases of two-bozon or two-fermion states). Some of such symmetrized given-configuration wave functions consist of two terms each, and in such cases the Schmidt entanglement can arise even if the sum in Eq. (\ref{dens-matr-multimode}) contains only one term. This is the case of the symmetry entanglement. In a general case of multiconfigurational states of indistinguishable particles, the Schmidt entanglement is determined by an unseparable superposition of the configurational and symmetry types of entanglement. Thus, the Schmidt decomposition (\ref{SCHMIDT-MODES}) and parameter $K$ (\ref{K}) characterize the total amount of entanglement in a pure bipartite state with both configurational and symmetry parts of entanglement completely taken into account.

A simple example of these definitions is the polarization state of two photons with different polarizations, horizontal ($H$) and vertical ($V$) ones, and with the state vector and the density-matrix operator given by
\begin{gather}
 %\nonumber
 |1_H,1_V\rangle=a_H^\dag a_V^\dag|0\rangle,\,
 \label{HV}
 \hat{\rho}=|1_H,1_V\rangle\langle 1_H,1_V|.
\end{gather}
Modes of this state are $H$ and $V$ and this is a single-configurational state with $n_H=1$ and  $n_V=1$. The state (\ref{HV}) does not have the configurational entanglement. For this reason it's often said that the state $|1_H,1_V\rangle$ is not entangled at all (see. e.g., Refs. \cite{Rubin,li}). But the discussion given above indicates clearly that this is not so because photons are indistinguishable bozon particles, and their symmetrized wave functions (\ref{conf-wf}) $\Psi_{HV}(\sigma_1,\sigma_2)$ and $\Psi_{HV}(\sigma_1^\prime,\sigma_2^\prime)$ for the only given configuration (\ref{HV}) consist of two terms each
\begin{equation}
 \label{conf-wf-HV}
 \Psi_{HV}(\sigma_1,\sigma_2)=\frac{\delta_{\sigma_1,H}\delta_{\sigma_2,V}
 +\delta_{\sigma_1,V}\delta_{\sigma_2,H}}{\sqrt{2}},
\end{equation}
and the same for $\Psi_{HV}^*(\sigma_1^\prime,\sigma_2^\prime)$. Here and below we use notations $\sigma_{1,2}$ for polarization variables instead of $x_{1,2}$ in Eqs. (\ref{SCHR}), (\ref{SCHMIDT-MODES}) and (\ref{dens-matr-multimode}), (\ref{conf-wf}). Note also, that, in contrast to distinguishable particles, in the case of indistinguishable particles (photons) the variables $\sigma_{1}$ and $\sigma_{1}$ are not associated with any specific particle, they have to be  interpreted as the polarization variables of one and another of two photons, never known specifically which one of them. Also, these variables are not associated with specific modes. E.g., as seen well from Eq. (\ref{conf-wf-HV}), the variable $\sigma_1$ can be related either to the mode $H$ or $V$. In this sense variables and modes are entangled, which means that there is no one-to-one correspondence between variables and modes. In accordance with Eq. (\ref{dens-matr-multimode}) and results of Ref. \cite{Archive}, symmetry of the wave function $\Psi_{HV}(\sigma_1,\sigma_2)$ (\ref{conf-wf-HV}) provides entanglement of the state $|1_H,1_V\rangle$ evaluated via the Schmidt decomposition and Schmidt parameter $K$, or concurrence $C$. This result agrees also with the conclusion of Ref. \cite{Pakauskas}.

However, the same state as given by Eq. (\ref{HV}) in the basis with horizontal and vertical axes $0x,0y$ ($0^\circ, 90^\circ$), in the basis turned for $45^\circ$ takes the form
\begin{gather}
 %\nonumber
 \frac{|2_{45^\circ},0_{135^\circ}\rangle-|0_{45^\circ},2_{135^\circ}\rangle}{\sqrt{2}}
 \label{45-135}
 =\frac{\big( a_{45^\circ}^{\dag\,2}-a_{135^\circ}^{\dag\,2}\big)|0\rangle}{2}.
\end{gather}
Two modes correspond now to photon polarizations along the $45^\circ$- and $135^\circ$-axes, and, obviously, the state (\ref{45-135}) is a two-configurational one. The configurations are ($n_{45^\circ}=2, n_{135^\circ}=0$) and ($n_{45^\circ}=0, n_{135^\circ}=2$). The sum over these two configurations in Eq. (\ref{dens-matr-multimode}) provides now the configurational entanglement. In the same time, the wave functions (\ref{conf-wf}) of the configurations  ($2_{45^\circ},0_{135^\circ}$) and ($0_{45^\circ},2_{135^\circ}$) are factorized, and in this case there is no symmetry entanglement.  Hence, in the basis ($45^\circ,135^\circ$) the Schmidt entanglement arises only owing to configurational entanglement, but both the degree of entanglement evaluated by the Schmidt parameter $K$, and the Schmidt modes remain the same as found in the basis ($0^\circ,90^\circ$). Thus, with changing bases, symmetry and configurational entanglement of pure polarization biphoton states are transformed into each other. In extreme cases of the bases ($0^\circ,90^\circ$) and ($45^\circ,135^\circ$) one of these two types of entanglement completely disappears. In all intermediate cases ($\alpha,\alpha+90^\circ$) both types of entanglement exist and have to be taken into account together. The Schmidt decomposition, Schmidt modes, and the Schmidt parameter $K$ characterize the total entanglement with both the symmetry and configurational contributions taken into account, and this makes the total entanglement basis-independent ($K(\alpha)=const.=2$). The total entanglement is the entanglement of a state as a whole, independent of conditions of its measurement or methods of its theoretical description.

One general remark more concerns spacial separation of indistinguishable particles, which is considered sometimes as partially diminishing their indistinguishability, making particles as if somewhat distinguishable. In the frame of such approach indistinguishable particles are assumed to get some features distinguishing them, e.g., as photons propagating to the left and to the right, with left- and right-propagating photons treated as usual distinguishable particles. Then, in accordance with Eq. (\ref{dens-matr-multimode}), only configurational entanglement is taken into account. In fact, such procedures are insufficient for finding total entanglement in systems of indistinguishable particles. Actually, any kind of a spacial separation creates a new degree of freedom for particles, but particles themselves remain indistinguishable. Bipartite states with several degrees of freedom (e.g., polarizations and propagation angles of photons or polarization and frequencies) are more complicated, and have a higher dimensionality than states with one degree of freedom (purely polarization biphoton states).

Polarization-frequency or polarization-angle ququarts are considered usually as two-qubit states, in which frequencies (or propagation angles) of photons are assumed to be strictly related to their variables, and symmetry of the wave function is not  taken into account \cite{Bogd}. In terms of discussed above configurational and symmetry entanglement, the two-qubit model of biphoton ququarts describes only the configurational entanglement and ignores the symmetry entanglement inevitably occurring in the case of indistinguishable particles. Though such approach can be reasonable in some cases, as said above, it is insufficient for evaluation of the total entanglement of biphoton ququarts. The approach appropriate for evaluation of the total entanglement was suggested in our work \cite{Archive}. In this approach biphoton polarization-frequency ququarts are considered as {\it two-qudit} states with the dimensionality of the one-photon Hilbert space $d=4$. Their wave function is symmetric with respect to the transposition of photons, and their entanglement is a combination of the polarization and frequency entanglement, as well as a combination of the configuration and symmetry entanglement. Features of frequency-polarization two-qudit biphoton ququarts are reminded briefly below in section {\bf II}. In section {\bf III} we discuss a representation of a general ququart's wave function in the form of sums of {\it double-Bell} states, where ``double-Bell$"$ means a product of, e.g., polarization and frequency Bell states. In sections {\bf IV} and {\bf V} we describe mixed biphoton states with ``invisible$"$ variables which arise from biphoton ququarts after averaging over either frequency or polarization variables. In experiment this corresponds to the situation when experimenters use detectors unsensitive either to values of frequencies of photons or to their polarizations, which makes these variable effectively invisible. Features of the arising two-frequency or two-polarization mixed states are rather unusual. We find explicitly their Schmidt parameter $K$, concurrence $C$, von Neumann mutual information $I$ and, in some simple cases, their relative entropy $C_{rel}$ and the degree of classical correlations $C_{cl}$. For biphoton mixed polarization states we find a universal relation between their degree of polarization and the Schmidt parameter $K$. On the other hand, we find that in the case of mixed polarization states the Schmidt parameter $K$ does not characterize anymore their degree of entanglement as it does in the case pure bipartite states. At last, in section {\bf VI} we discuss in more details a relationship between features of polarization biphoton ququarts with invisible frequency variables and of two-qubit ququarts with  asymmetric wave functions and photon frequencies not considered as variables at all. We discuss experimental schemes in which biphoton polarization-frequency ququarts can display either features of full-dimensionality two-dudit states, or of a two-qubit model, or of mixed polarization states averaged over frequencies.

\section{Biphoton ququarts}
Biphoton ququarts can be produced, e.g., in collinear non-degenerate SPDC processes [two possible frequencies, $\omega_h$ (high, $h$) or $\omega_l$ (low, $l$) and one of two polarizations for each photon, horizonal ($H$) and vertical ($V$)]. To get ququarts of a general form, on can use more than one nonlinear birefringent crystal and/or additional manipulations with photon polarizations after crystals. Given frequencies of photons in SPDC pairs ($\omega_h$ and $\omega_l$) can be obtained either with the help of frequency filters or in the approximation of a highly monochromatic pump and a very long crystal. In such schemes, polarizations and frequencies of photons are two independent degrees of freedom, and the corresponding variables $\sigma_i$ and $\omega_i$ ($i=1,2$) can take two values each for each photon, $\sigma_i=H$ or $V$ and $\omega_i=\omega_h$ or $\omega_l$. The only restriction is that frequencies of two photons are always different, $\omega_1\neq\omega_2$. One-photon polarization-frequency basis modes are $Hh,\,Hl,\,Vh,\,{\rm and}\,Vl$. The corresponding  state-vectors and wave functions are given by
\begin{equation}
 \label{qudit}
 \begin{matrix}
 {\displaystyle{\rm state}\atop\displaystyle{\rm vector}} & {\rm wave\, function}\\
 a_{Hh}^\dag|0\rangle &
 \delta_{\sigma,\,H}\delta_{\omega,\,\omega_h}=\left(\begin{matrix}1\\0\end{matrix}\right)^{pol}\otimes
 \left(\begin{matrix}1\\0\end{matrix}\right)^{freq}
 =\left(\scriptsize{\begin{matrix}1\\0\\0\\0\end{matrix}}\right) ,\\
 a_{Hl}^\dag|0\rangle &
 \delta_{\sigma,\,H}\delta_{\omega,\,\omega_l}
 =\left(\begin{matrix}1\\0\end{matrix}\right)^{pol}\otimes
 \left(\begin{matrix}0\\1\end{matrix}\right)^{freq}
 =\left(\scriptsize{\begin{matrix}0\\1\\0\\0\end{matrix}}\right),\\
 a_{Vh}^\dag|0\rangle &
 \delta_{\sigma,\,V}\delta_{\omega,\,\omega_h}
 =\left(\begin{matrix}0\\1\end{matrix}\right)^{pol}\otimes
 \left(\begin{matrix}1\\0\end{matrix}\right)^{freq}
 =\left(\scriptsize{\begin{matrix}0\\0\\1\\0\end{matrix}}\right),\\
 a_{Vl}^\dag|0\rangle &
  \delta_{\sigma,\,V}\delta_{\omega,\,\omega_l}=
\left(\begin{matrix}0\\1\end{matrix}\right)^{pol}\otimes
 \left(\begin{matrix}0\\1\end{matrix}\right)^{freq}
 =\left(\scriptsize{\begin{matrix}0\\0\\0\\1\end{matrix}}\right).
 \end{matrix}
\end{equation}
Here the superscripts $pol$ and $freq$ refer the polarization and frequency degrees of freedom. The upper and lower lines in the two-line columns correspond to the horizonal polarization and high frequency, whereas the lower lines - to the vertical polarization and low frequency. Altogether, four one-photon states (\ref{qudit}) form a qudit with the dimensionality of the one-photon Hilbert space $d=4$.

Under the assumption that frequencies of two photons in the SPDC pairs are always different one can construct from four one-photon states (\ref{qudit}) only four independent basis biphoton states with the state vectors
\begin{gather}
 \nonumber
 |\Psi_{HH}^{(4)}\rangle=\,a_{Hh}^\dag a_{Hl}^\dag|0\rangle,\,
 |\Psi_{VV}^{(4)}\rangle=\,a_{Vh}^\dag a_{Vl}^\dag|0\rangle,\\
 \nonumber
 |\Psi_{HV}^{(4)}\rangle=\,a_{Hh}^\dag a_{Vl}^\dag|0\rangle,\\
 \label{two-phot-bas-st-vec}
 |\Psi_{VH}^{(4)}\rangle=\,a_{Vh}^\dag a_{Hl}^\dag|0\rangle.
\end{gather}
and corresponding to them wave functions \cite{Archive}
\begin{gather}
 \nonumber
 \Psi^{(4)}_{HH}=\frac{1}{\sqrt{2}}\scriptsize{\left(\begin{matrix}1\\0\end{matrix}\right)_1^{pol}
 \otimes\left(\begin{matrix}1\\0\end{matrix}\right)_2^{pol}}\\
 \nonumber
 \scriptsize{\otimes\left[\left(\begin{matrix}1\\0\end{matrix}\right)_1\otimes\left(\begin{matrix}0\\ 1\end{matrix}\right)_2
 +\left(\begin{matrix}0\\1\end{matrix}\right)_1\otimes\left(\begin{matrix}1\\0\end{matrix}\right)_2\right]^{freq}}\\
 \label{HH-4}
 \equiv\frac{1}{\sqrt{2}}\left\{\scriptsize{
 \left(\begin{matrix}1\\0\\0\\0\end{matrix}\right)_1\otimes\left(\begin{matrix}0\\1\\0\\ 0\end{matrix}\right)_2+\left(\begin{matrix}0\\1\\0\\0\end{matrix}\right)_1
 \otimes\left(\begin{matrix}1\\0\\0\\0\end{matrix}\right)_2}\right\},
\end{gather}

\begin{gather}
 \nonumber
 \Psi^{(4)}_{HV}=\frac{1}{\sqrt{2}}\scriptsize{\left[\left(\begin{matrix}1\\0\end{matrix}\right)_1^{pol}
 \otimes\left(\begin{matrix}0\\1\end{matrix}\right)_2^{pol}\otimes\left(\begin{matrix}1\\ 0\end{matrix}\right)_1^{frec}\otimes\left(\begin{matrix}0\\1\end{matrix}\right)_2^{frec}\right.}\\
 \nonumber
 +\scriptsize{\left.\left(\begin{matrix}0\\1\end{matrix}\right)_1^{pol}\otimes\left(\begin{matrix}1\\ 0\end{matrix}\right)_2^{pol}\otimes
 \left(\begin{matrix}0\\1\end{matrix}\right)_1^{frec}\otimes\left(\begin{matrix}1\\0\end{matrix}\right)_2^{frec}
 \right]}\\
 \label{HV-4}
 \equiv\frac{1}{\sqrt{2}}\scriptsize{\left\{
 \left(\begin{matrix}1\\0\\0\\0\end{matrix}\right)_1\otimes\left(\begin{matrix}0\\0\\0\\1\end{matrix}\right)_2
 +\left(\begin{matrix}0\\0\\0\\1\end{matrix}\right)_1\otimes\left(\begin{matrix}1\\0\\0\\0\end{matrix}\right)_2
 \right\}},
\end{gather}

\begin{gather}
 \nonumber
 \Psi^{(4)}_{VH}=\frac{1}{\sqrt{2}}\scriptsize{\left[\left(\begin{matrix}0\\1\end{matrix}\right)_1^{pol}
 \otimes\left(\begin{matrix}1\\0\end{matrix}\right)_2^{pol}\otimes\left(\begin{matrix}1\\ 0\end{matrix}\right)_1^{frec}\otimes\left(\begin{matrix}0\\1\end{matrix}\right)_2^{frec}\right.}\\
 \nonumber
 \scriptsize{+\left.\left(\begin{matrix}1\\0\end{matrix}\right)_1^{pol}\otimes
 \left(\begin{matrix}0\\1\end{matrix}\right)_2^{pol}
 \left(\begin{matrix}0\\1\end{matrix}\right)_1^{frec}
 \otimes\left(\begin{matrix}1\\0\end{matrix}\right)_2^{frec}\right]}\\
 \label{VH-4}
 \equiv\frac{1}{\sqrt{2}}\scriptsize{\left\{
 \left(\begin{matrix}0\\0\\1\\0\end{matrix}\right)_1\otimes
 \left(\begin{matrix}0\\1\\0\\0\end{matrix}\right)_2+\left(\begin{matrix}0\\1\\0\\ 0\end{matrix}\right)_1\otimes\left(\begin{matrix}0\\0\\1\\0\end{matrix}\right)_2
 \right\}},
\end{gather}

\begin{gather}
 \nonumber
 \Psi^{(4)}_{VV}=\scriptsize{\left(\begin{matrix}0\\1\end{matrix}\right)_1^{pol}
 \otimes\left(\begin{matrix}0\\1\end{matrix}\right)_2^{pol}}\\
 \nonumber
 \otimes\frac{1}{\sqrt{2}}\scriptsize{\left[\left(\begin{matrix}1\\0\end{matrix}\right)_1\otimes
 \left(\begin{matrix}0\\1\end{matrix}\right)_2+\left(\begin{matrix}0\\1\end{matrix}\right)_1
 \otimes\left(\begin{matrix}1\\0\end{matrix}\right)_2\right]^{frec}}\\
 \label{VV-4}
 \equiv\scriptsize{\frac{1}{\sqrt{2}}\left\{
 \left(\begin{matrix}0\\0\\1\\0\end{matrix}\right)_1
 \otimes\left(\begin{matrix}0\\0\\0\\1\end{matrix}\right)_2
 +\left(\begin{matrix}0\\0\\0\\1\end{matrix}\right)_1
 \otimes\left(\begin{matrix}0\\0\\1\\0\end{matrix}\right)_2
 \right\}}.
\end{gather}
where the indices 1 and 2 numerate variables of indistinguishable photons $\sigma_1,\omega_1$ and $\sigma_2,\omega_2$ (not seen explicitly in the matrix form of writing the wave functions). All basis wave functions (\ref{HH-4})-(\ref{VV-4}) are symmetric with respect to the transposition of particle variables $1\rightleftharpoons 2$, as it should be for any two-bozon states. In a general form, the wave function of a ququart is defined as a superposition of four basis wave function (\ref{HH-4})-(\ref{VV-4})
\begin{equation}
 \label{ququart}
 \Psi^{(4)}=C_1\Psi^{(4)}_{HH}+C_2\Psi^{(4)}_{HV}+C_3\Psi^{(4)}_{VH}+C_4\Psi^{(4)}_{VV}
\end{equation}
with arbitrary coefficients $C_{1,2,3,4}$ obeying the normalization condition $|C_1|^2+|C_2|^2+|C_3|^2+|C_4|^2=1$.

Note that in Eqs. (\ref{HH-4})-(\ref{VV-4}) the two-row columns are combined into four-row ones corresponding to given particle variables but mixed up polarization and frequency degrees of freedom. Alternatively, it's possible to combine parts of wave functions with given degrees of freedom but mixed up variables to rewrite Eqs. (\ref{HH-4})-(\ref{VV-4}) as
\begin{gather}
 \label{HH-altern}
 \Psi_{HH}^{(4)}=\scriptsize{  \left(\begin{matrix}1\\0\\0\\0\end{matrix}\right)^{pol}\otimes
  \left(\begin{matrix}0\\1/\sqrt{2}\\1/\sqrt{2}\\0\end{matrix}\right)^{freq}},\\
  \label{HV-altern}
  \Psi_{HV}^{(4)}=\frac{1}{\sqrt{2}}\left\{\scriptsize{  \left(\begin{matrix}0\\1\\0\\0\end{matrix}\right)^{pol}\otimes
  \left(\begin{matrix}0\\1\\0\\0\end{matrix}\right)^{freq}+
  \left(\begin{matrix}0\\0\\1\\0\end{matrix}\right)^{pol}\otimes
  \left(\begin{matrix}0\\0\\1\\0\end{matrix}\right)^{freq}}\right\},\\
  \label{VH-altern}
  \Psi_{HV}^{(4)}=\frac{1}{\sqrt{2}}\left\{\scriptsize{  \left(\begin{matrix}0\\0\\1\\0\end{matrix}\right)^{pol}\otimes
  \left(\begin{matrix}0\\1\\0\\0\end{matrix}\right)^{freq}+
  \left(\begin{matrix}0\\1\\0\\0\end{matrix}\right)^{pol}\otimes
  \left(\begin{matrix}0\\0\\1\\0\end{matrix}\right)^{freq}}\right\},\\
  \label{VV-altern}
 \Psi_{VV}^{(4)}=\scriptsize{  \left(\begin{matrix}0\\0\\0\\1\end{matrix}\right)^{pol}\otimes
  \left(\begin{matrix}0\\1/\sqrt{2}\\1/\sqrt{2}\\0\end{matrix}\right)^{freq}}.
\end{gather}
Biphoton basis states written in the form of Eqs. (\ref{HH-4})-(\ref{VV-4}) are convenient for reduction of the density matrix over one of two pairs of photon variables, ($\sigma_1,\, \omega_1$) or ($\sigma_2,\, \omega_2$), whereas the form (\ref{HH-altern})-(\ref{VV-altern}) is more convenient for reduction in one of two degrees of freedom, polarization or frequency, i.e. for transition to mixed states.

As evident from Eqs. (\ref{HH-4})-(\ref{VV-4})  and (\ref{HH-altern})-(\ref{VV-altern}), biphoton ququarts  describe two-qudit states of dimensionality $D=d^2=16$, and they can be considered as two-qudit states (with $d=4$) \cite{Archive}. This understanding differs from a widely spread opinion that biphoton ququarts are two-qubit states. The two-qubit model is correct only for ququarts of two distinguishable particles with only one degree of freedom taken into account. In contrast to this, in the case of biphoton ququarts we have two equally important degrees of freedom (e.g., polarization and frequency) and, owing to obligatory symmetry of two-bozon states, entanglement in both degrees of freedom is possible and has to be taken into account. This makes features of pure-state biphoton ququarts rather significantly different from those of the two-qubit ququarts \cite{Archive}.

Note also that instead of frequencies $\omega_h$ and $\omega_l$ one can consider photons with coinciding frequencies but different directions of propagation (in the non-collinear SPDC scheme) characterized by angles $\theta_{left}$ and $\theta_{right}$. In this case two degrees of freedom  are polarizations and angles, determining directions of propagation of photons, and two sets of variables are ($\sigma_1,\theta_1$) and ($\sigma_2,\theta_2$). Except different notations, the physics of polarization-frequency and polarization-angle biphoton ququarts is identical.

\section {Double-Bell states}

Instead of four basis wave functions of ququarts (\ref{HH-4})-(\ref{VV-4}) one can use their superpositions:
\begin{equation}
 \label{superp-basis}
 \setlength{\extrarowheight}{0.3cm}
 \begin{matrix}
 \Phi_{\pm}^{(4)}=\frac{1}{\sqrt{2}}\left(\Psi^{(4)}_{HH}\pm\Psi^{(4)}_{VV}\right),\\
 \Psi_{\pm}^{(4)}=\frac{1}{\sqrt{2}}\left(\Psi^{(4)}_{HV}\pm\Psi^{(4)}_{VH}\right).
 \end{matrix}
\end{equation}
The sum of these basis functions with arbitrary coefficients gives a general expression for the  wave function of a biphoton quaquart equivalent to that of Eq. (\ref{ququart})
\begin{equation}
 \label{qqrt-dB-exp}
 \Psi^{(4)}=C_+\Phi_+^{(4)}+B_+\Psi_+^{(4)}+B_-\Psi_-^{(4)}+C_-\Phi_-^{(4)},
\end{equation}
where $C_\pm=\left(C_1\pm C_4\right)/\sqrt{2}$ and $B_\pm=\left(C_2\pm C_3\right)/\sqrt{2}$.

Below we will use also a mixture of basis wave functions (\ref{HH-4})-(\ref{VV-4}) and (\ref{superp-basis}):
\begin{equation}
 \label{mixed basis}
 \Big\{\Psi^{(4)}_{HH},\;\Psi_{+}^{(4)},\;\Psi^{(4)}_{VV},\;\Psi_{-}^{(4)}\Big\}.
\end{equation}
Expansion of a general ququart's wave function in the basis functions of Eq. (\ref{mixed basis}) is given by
\begin{equation}
 \label{ququart-half-Bell}
 \Psi^{(4)}=C_1\Psi^{(4)}_{HH}+B_+\Psi_+^{(4)}+C_4\Psi^{(4)}_{VV}+B_-\Psi_-^{(4)},
\end{equation}
with
\begin{equation}
 \label{norm}
 |C_1|^2+|B_+|^2|+|B_-|^2+|C_4|^2=1.
\end{equation}

The wave functions $\Psi^{(4)}_{\pm}$ and $\Phi_{\pm}^{(4)}$ of Eq. (\ref{superp-basis}) are not Bell states because they have a higher dimensionality ($D=16$) than the 4-dimensional two-qubit true Bell states
\begin{equation}
 \label{Bell-Psi}
  \Psi_\pm^{(Bell)}=\frac{1}{\sqrt{2}}{\scriptsize \left[\left(\begin{matrix}1\\0\end{matrix}\right)_1
 \otimes\left(\begin{matrix}0\\1\end{matrix}\right)_2\pm\left(\begin{matrix}0\\1\end{matrix}\right)_1
 \otimes\left(\begin{matrix}1\\0\end{matrix}\right)_2\right]\equiv\frac{1}{\sqrt{2}}\left(\begin{matrix}0\\1\\ \pm 1\\0\end{matrix}\right)},
\end{equation}
\begin{equation}
   \label{Bell-Phi}
    \Phi_\pm^{(Bell)}=
    \frac{1}{\sqrt{2}}{\scriptsize\left[\left(\begin{matrix}1\\0\end{matrix}\right)_1
    \otimes\left(\begin{matrix}1\\0\end{matrix}\right)_2\pm
    \left(\begin{matrix}0\\1\end{matrix}\right)_1
    \otimes\left(\begin{matrix}0\\1\end{matrix}\right)_2\right]
    \equiv\frac{1}{\sqrt{2}}\left(\begin{matrix}1\\0\\0\\ \pm 1\end{matrix}\right).}
\end{equation}
But, as can be easily checked, the functions $\Psi^{(4)}_{\pm}$ and $\Phi_{\pm}^{(4)}$ can be presented as direct products of polarization and frequency Bell states, and these products can be referred to as double-Bell states:
\begin{equation}
 \label{Double-Bell}
 \setlength{\extrarowheight}{0.3cm}
 \begin{matrix}
 \Phi^{(4)}_{\pm}=\Phi_\pm^{(Bell)\,pol}\otimes\Psi_+^{(Bell)\,freq},\\
 \Psi^{(4)}_{\pm}=\Psi_\pm^{(Bell)\,pol}\otimes\Psi_\pm^{(Bell)\,freq}.
 \end{matrix}
\end{equation}
With these equalities taken into account, the general wave function of a biphoton ququart (\ref{ququart-half-Bell}) can be rewritten as
\begin{equation}
 \label{ququart-dB}
 \Psi^{(4)}=\Psi^{(3)\,pol}\otimes\Psi_+^{(Bell)\,freq}+ B_-\Psi_-^{(Bell)\,pol}\otimes\Psi_-^{(Bell)\,freq},
\end{equation}
where
\begin{equation}
\label{qtrt-wf}
 \Psi^{(3)\,pol}=C_1\Psi^{pol}_{HH}+B_+\Psi_+^{(Bell)\,pol}+C_4\Psi^{pol}_{VV}
\end{equation}
is a general wave function of a polarization biphoton qutrit \cite{Archive};
\begin{equation}
 \label{HH-3 and VV-3}
 \Psi^{pol}_{HH}={\scriptsize\left(\begin{matrix}1\\0\end{matrix}\right)_1^{pol}\otimes
 \left(\begin{matrix}1\\0\end{matrix}\right)_2^{pol},\;  \Psi^{pol}_{VV}=
 \left(\begin{matrix}0\\1\end{matrix}\right)_1^{pol}\otimes
 \left(\begin{matrix}0\\1\end{matrix}\right)_2^{pol}}
\end{equation}
are purely polarization states of two photons with coinciding polarizations.
For comparison with notations of Ref. \cite{Archive}, $B_+$ and $C_4$ in Eq. (\ref{qtrt-wf}) correspond to $C_2$ and $C_3$ for qutrits in \cite{Archive}.

The density matrix of a ququart (\ref{ququart-dB}) is given by
\begin{gather}
 \nonumber
 \rho^{(4)}=\\
 \nonumber
 \left[\Psi^{(3)\,pol}\otimes\Psi_+^{(Bell)\,freq}+B_-\Psi_-^{(Bell)\,pol}\otimes\Psi_-^{(Bell)\,freq}\right]\\
 \label{rho-ququart-dB}
 \otimes\left[\Psi^{(3)\,pol}\otimes\Psi_+^{(Bell)\,freq}
 +B_-\Psi_-^{(Bell)\,pol}\otimes\Psi_-^{(Bell)\,freq}\right]^\dag.
\end{gather}

As mentioned above, the dimensionality of the density matrix $\rho^{(4)}$ is $16\times 16$. But the number of ququart's basis states in all equations (\ref{two-phot-bas-st-vec})-(\ref{ququart-dB}) is only 4 rather than 16. This means in fact, that 12 missing basis states come in the superpositions forming the biphoton ququarts with zero coefficients. The excluded states are states with a wrong symmetry and states with coinciding frequencies of photons. Owing to these omissions the $16\times 16$ density matrix has many zero elements. Positions of these zero elements depend on a choice of a basis used for calculation of the density matrix. The  density matrix has the simplest form if it is calculated in the basis of ququart's basis states, e.g., in the basis of states (\ref{mixed basis}), $\Big\{\Psi^{(4)}_{HH},\;\Psi_{+}^{(4)},\;\Psi_{-}^{(4)},\;\Psi^{(4)}_{VV}\Big\}$ plus 12 other unspecified basis states coming with zero coefficients in the ququart's wave function (\ref{ququart-half-Bell}). With such choice of the basis for calculation of the density matrix, its 4 nonzero lines and columns can be concentrated in the upper left corner. Then, the nonzero $4\times 4$ part of the density matrix (\ref{rho-ququart-dB}) can be written as
\begin{equation}
 \label{qqrt-dm-4x4}
 \rho^{(4)}=\left(
 \setlength{\extrarowheight}{0.3cm}
 \begin{matrix}
 |C_1|^2 & C_1B_+^* & C_1C_4^* & C_1B_-^*\\
 B_+C_1^* & |B_+|^2 & B_+C_4^* & B_+B_-^*\\
 C_4C_1^* & C_4B_+^* & |C_4|^2 & C_4B_-^*\\
 B_-C_1^* & B_-B_+^* & B_-C_4^* & |B_-|^2
 \end{matrix}
 \right).
\end{equation}

\section{Biphoton ququarts with ``invisible$"$ variables}

For measuring parameters of biphoton ququarts one has to use detectors (photon counters) provided with polarizers and frequency filters to distinguish photons in all four polarization-frequency modes (\ref{qudit}) \cite{Archive}. However, it's interesting also what happens if one uses only part of these devices, either only polarizers or only frequency filters. In these two cases either frequency or polarization variables become ``invisible$"$. Mathematically, such states are characterized by density matrices $\rho^{pol}=Tr_{freq}\rho$ or $\rho^{freq}=Tr_{pol}\rho$, where $\rho$ is the general density matrix of Eqs. (\ref{rho-ququart-dB}), (\ref{qqrt-dm-4x4}), and $Tr_{freq}$, $Tr_{pol}$ denote partial traces over frequency or polarization variables. Features of such states with ``invisible$"$ variables are significantly different from those of both original biphoton ququarts and purely polarization or frequency biphoton states. If original biphoton ququarts are pure states, biphoton ququarts with invisible variable are, typically, mixed. Besides, if density matrix of the original biphoton ququart is 16-dimensional, the matrices  $\rho^{pol}$ and $\rho^{freq}$ have the dimensionality $D=4$ and describe in a general case two-qubit mixed states. Features of such states are rather interesting and they are described below. As the first example, let us consider the two-frequency states with ``invisible$"$ polarization variables.

\subsection*{Two-frequency state with ``invisible$"$ polarization variables}

The density matrix $\rho^{freq}=Tr_{pol}\rho$ is easily found from  Eq. (\ref{rho-ququart-dB}) to be given by
\begin{gather}
 \nonumber
 \rho^{freq}=(1-|B_-|^2)\Psi_+^{(Bell)\, freq}\otimes\left(\Psi_+^{(Bell)\, freq}\right)^\dag\\
 \label{freq-mixed-rho}
 +|B_-|^2\Psi_-^{(Bell)\, freq}\otimes\left(\Psi_-^{(Bell)\, freq}\right)^\dag\\
 \label{freq-4x4}
 =\frac{1}{2}\left(
 \begin{matrix}
 0 & 0 & 0 & 0\\
 0 & 1 & 1-2|B_-|^2 & 0\\
 0 & \;1-2|B_-|^2 & 1 & 0\\
 0 & 0 & 0 & 0
 \end{matrix}
 \right)_{nat.\,basis}\\
 \label{2times2}
 =
 \left(
 \begin{matrix}
 1-|B_-|^2 & 0\\
 0 & |B_-|^2
 \end{matrix}
 \right)_{Bell\,basis}.
\end{gather}
Eq. (\ref{freq-4x4}) corresponds to writing the frequency density matrix in the natural basis ${\tiny\left\{\left(\begin{matrix}1\\ 0\\0\\0\end{matrix}\right),\left(\begin{matrix}0\\1\\0\\0\end{matrix}\right),\left(\begin{matrix}0\\0\\1\\ 0\end{matrix}\right),\left(\begin{matrix}0\\0\\0\\1\end{matrix}\right)\right\}}$, whereas Eq. (\ref{2times2}) - to its presentation in the basis of Bell states $\left\{\Psi_+^{(Bell)\, freq},\Psi_-^{(Bell)\, freq},\Phi_+^{(Bell)\, freq},\Phi_+^{(Bell)\, freq}\right\}$ with zero lines and columns dropped.

\begin{figure}[h]
\centering\includegraphics[width=7cm]{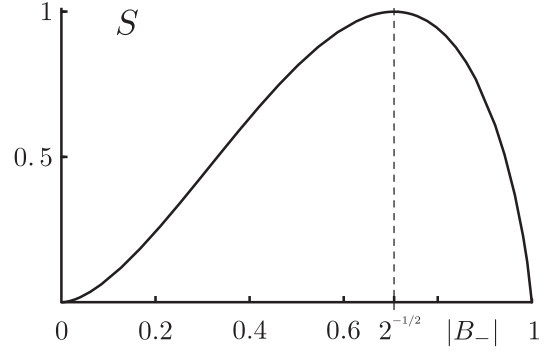}
\caption{{\protect\footnotesize {Entropy (\ref{S(rho-freq)}) of the two-frequency mixed state with the density matrix $\rho^{freq}$ (\ref{freq-4x4}), (\ref{freq-2x2x2}) as a function of $|B_-|$}}}\label{Fig1}
\end{figure}

As it's evident from Eqs. (\ref{freq-4x4}) and (\ref{2times2}), eigenvalues and entropy of the matrix $\rho^{freq}$ are equal to
\begin{equation}
 \label{eigenvalues-full}
 \lambda_1=\lambda_2=0,\; \lambda_{3}=1-|B_-|^2,\;\lambda_{4}=|B_-|^2
\end{equation}
and
\begin{gather}
 \nonumber
 S(\rho^{freq})=-\sum_i\lambda_i\log_2\lambda_i= -2|B_-|^2\log_2|B_-|\\
 \label{S(rho-freq)}
-(1-|B_-|^2)\log_2(1-|B_-|^2).
\end{gather}
The entropy $S(\rho^{freq})$ is shown in Fig. \ref{Fig1} as a function of $|B_-|$.
It's seen to turn zero only at $B_-=0$ and $|B_-|=1$. In these two cases the density matrix $\rho^{freq}$ describes pure states: at $B_-=0$ Eq. (\ref{freq-mixed-rho}) yields $\rho^{freq}=\Psi_+^{(Bell)\, freq}\otimes\left(\Psi_+^{(Bell)\, freq}\right)^\dag$, and at $|B_-|=1$ $\rho^{freq}=\Psi_-^{(Bell)\, freq}\otimes\left(\Psi_-^{(Bell)\, freq}\right)^\dag$. In all other cases, $|B_-|\neq 0, 1$, the state characterized by the density matrix $\rho^{freq}$ (\ref{freq-mixed-rho})-(\ref{2times2}) is mixed.

To calculate the reduced density matrix of the mixed two-frequency state, we have to present the full matrix $\rho^{freq}$ (\ref{freq-mixed-rho}) in the form of a sum of products of $2\times 2$ matrices
\begin{gather}
 \nonumber
 \rho^{freq}=\sum_{\pm}\frac{1\pm(1-2|B_-|^2)}{4}\\
 \nonumber
 \times\left\{\left(\begin{matrix}1&0\\0&0\end{matrix}\right)_1\otimes\left(\begin{matrix}0&0\\0&1\end{matrix}
 \right)_2\pm\left(\begin{matrix}0&1\\0&0\end{matrix}\right)_1\otimes\left(\begin{matrix}0&0\\1&0\end{matrix}
 \right)_2\right.\\
 \pm\left(\begin{matrix}0&0\\1&0\end{matrix}\right)_1\otimes\left(\begin{matrix}0&1\\0&0\end{matrix}
 \right)
 \label{freq-2x2x2}
 +\left.\left(\begin{matrix}0&0\\0&1\end{matrix}\right)_1\otimes\left(\begin{matrix}1&0\\0&0\end{matrix}
 \right)_2
 \right\}.
\end{gather}
By taking the trace of this matrix, e.g., with respect to the frequency variable of a photon 2, we easily find that the reduced density matrix is given by
\begin{equation}
 \label{freq-reduced}
 \rho_r^{freq}=\frac{1}{2}
 \left(\begin{matrix} 1 & 0 \\0 & 1\end{matrix}\right).
\end{equation}
Eigenvalues of this matrix are are $\lambda_1^{(r)}=\lambda_2^{(r)}=\frac{1}{2}$, which corresponds to the Schmidt parameter $K^{freq}=[\lambda_1^{(r)\,2}+\lambda_2^{(r)\,2}]^{-1}=2$ and entropy of the reduced state $S(\rho_r^{freq})=-2\times\frac{1}{2}\times\log_2\frac{1}{2}=1$. By combining the last result with that of Eq. (\ref{S(rho-freq)}), we find the von-Neumann mutual information of the mixed frequency state characterized by the density matrix of Eqs. (\ref{freq-mixed-rho}), (\ref{freq-4x4})
\begin{gather}
 \nonumber
 I(\rho^{freq})=2S(\rho_r^{freq})-S(\rho^{freq}) =2-S(\rho^{freq})\\
 \label{I(rho-freq)}
 =2+|B_-|^2\log_2|B_-|^2+(1-|B_-|^2)\log_2(1-|B_-|^2)
\end{gather}
in agreement with the results of Ref. \cite{Henderson}.
The concurrence of the mixed state under consideration can be easily found directly from the first expression of Eq. (\ref{freq-4x4}) with the help of the Wootters' procedure \cite{Wootters}, and it's given by
\begin{equation}
 \label{conc-freq}
 C^{freq}=|1-2|B_-|^2|.
\end{equation}
In addition to the concurrence, there are several other measures of entanglement valid for mixed states. One of them is the relative entropy suggested by Vedral et al \cite{Vedral-Knight, Vedral, Henderson}, which is defined as the ``distance$"$ between the density matrix $\rho$ and the closest unentangled density matrix $\sigma$: $S_{rel}(\rho)=Tr[\rho(\log_2\rho-\log_2\sigma)]$. For the state of the form (\ref{freq-mixed-rho}) the relative entropy is known \cite{Henderson} and in a slightly modified form it can be presented as  (\ref{freq-mixed-rho})
\begin{gather}
 \nonumber
 S_{rel}(\rho^{freq})=\frac{1+C^{freq}}{2}\log_2(1+C^{freq})\\
 \label{S(rel-freq)}
 +\frac{1-C^{freq}}{2}\log_2(1-C^{freq}).
\end{gather}

\begin{figure}[h]
\centering\includegraphics[width=7cm]{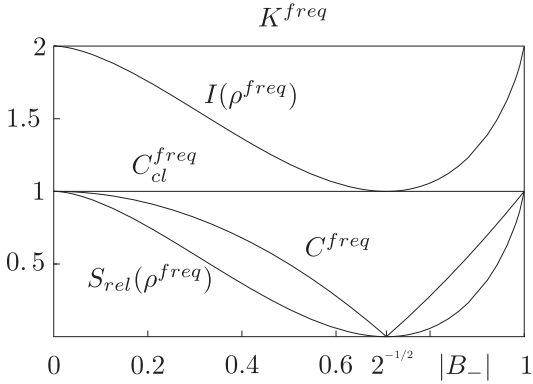}
\caption{{\protect\footnotesize {Parameters, characterizing degrees of correlations in the two-frequency mixed state with the density matrix $\rho^{freq}$ (\ref{freq-4x4}) as functions of $|B_-|$}.}}\label{Fig2}
\end{figure}

At last, one possibility of quantifying the amount of classical correlations is related to the definition of works \cite{Henderson, Hamieh, Oh-Kim} where the degree of classical correlations was defined as the difference between the von Neumann mutual information and the relative entropy, for the state (\ref{freq-mixed-rho}) given by
\begin{equation}
 \label{Class-freq}
 C_{cl}^{freq}=I(\rho^{freq})-S_{rel}(\rho^{freq})\equiv 1,
\end{equation}
again, in agreement with the conclusion of Ref. \cite{Henderson}.
All characteristics of correlations in the mixed two-frequency state are shown in Fig. \ref{Fig2} as functions of the parameter $|B_-|$.
This picture shows that: 1) The relative entropy is always less than the concurrence except the cases of pure states ($B_-=0$ or 1) when they are equal; hence, in the case of mixed states the relative entropy better characterizes the level of quantum correlations than concurrence. 2) In the case of pure states entropic characteristics of the levels of classical and quantum correlations are equal, and each of them equals a half of the von Neumann mutual information, $C_{cl}=C^{freq}=S_r=\frac{1}{2}I(\rho^{freq})$ in the cases $B_-=0$ and $|B_-|=1$. 3) The Schmidt parameter $K$ is a non-entropic characteristic which can be used for evaluation of the levels of both classical and quantum correlations only in the case of pure states; in the case of mixed states $K(|B_-|)$ behaves as $C_{cl}(|B_-|)$ (they are both constant) and, hence, for mixed states, $K$ can be considered only as a non-entropic measure of classical correlations; relation between the Schmidt parameter and concurrence $C=\sqrt{2(1-K^{-1})}$ occurs only in the case of pure states and is inapplicable for mixed states.

\section{Mixed polarization biphoton states with ``invisible$"$ frequency variables}
If we decide to choose the frequency variables of photons as ``invisible" ones, we have to take a trace of the general density matrix $\rho^{(4)}$ (\ref{rho-ququart-dB}) with respect to ``invisible" frequency variables to get the density matrix of a mixed two-photon polarization state
\begin{equation}
 \label{rho-pol-mixed-general}
 \rho^{pol}=\Psi^{(3)\,pol}\otimes\left(\Psi^{(3)\,pol}\right)^\dag
 +|B_-|^2\Psi_-^{(Bell) pol}\otimes\left(\Psi_-^{(Bell) pol}\right)^\dag.
 \end{equation}
 \subsection{Reduced density matrix, eigenvalues, Schmidt paramerter, and entropy}
For finding further the reduced density matrix of the mixed polarization biophoton state, let us write down first the density matrix $\rho^{pol}$ in the form of products of single-particle matrices
\begin{gather}
 \nonumber
 \rho^{pol}=|C_1|^2\left(\begin{matrix}1&0\\0&0\end{matrix}\right)_1\otimes
 \left(\begin{matrix}1&0\\0&0\end{matrix}\right)_2+
 \frac{|B_+|^2+|B_-|^2}{2}\\
 \nonumber
 \times\left[\left(\begin{matrix}1&0\\0&0\end{matrix}\right)_1\otimes\left(\begin{matrix}0&0\\0&1\end{matrix}
 \right)_2+
 \left(\begin{matrix}0&0\\0&1\end{matrix}\right)_1\otimes\left(\begin{matrix}1&0\\0&0\end{matrix}
 \right)_2\right]\\
 \nonumber
 +|C_4|^2\left(\begin{matrix}0&0\\0&1\end{matrix}\right)_1\otimes\left(\begin{matrix}0&0\\0&1\end{matrix}
 \right)_2\\
 \nonumber
 +\frac{1}{\sqrt{2}}
 \left[C_1B_+^*\left(\begin{matrix}0&1\\0&0\end{matrix}\right)_1+
 C_1^*B_+\left(\begin{matrix}0&0\\1&0\end{matrix}\right)_1\,\right]
 \otimes\left(\begin{matrix}1&0\\0&0\end{matrix}\right)_2\\
 \label{pol-2x2x2}
 +\frac{1}{\sqrt{2}}\left[B_+C_4^*\left(\begin{matrix}0&1\\0&0\end{matrix}\right)_1
 +B_+^*C_4\left(\begin{matrix}0&0\\1&0\end{matrix}\right)_1\,\right]
 \otimes\left(\begin{matrix}0&0\\0&1\end{matrix}\right)_2 + ... .
\end{gather}
In this expression only a half of terms is written explicitly whereas the other part is indicated by ... and, actually, dropped as unimportant for calculation of the density matrix reduced over the variable 2, because in these terms all matrices $\left(*\;*\atop{*\;*}\right)_2$ are off-diagonal. Thus, the reduced density matrix determined by Eq. (\ref{pol-2x2x2}) is given by
\begin{gather}
 \nonumber
 \rho^{pol}_r=\left(
 \begin{matrix}
|C_1|^2+\frac{|B_+|^2+|B_-|^2}{2} & \frac{C_1B_+^*+B_+C_4^*}{\sqrt{2}}\\
 \frac{C_1^*B_++B_+^*C_4^*}{\sqrt{2}} & |C_4|^2+\frac{|B_+|^2+|B_-|^2}{2}
 \end{matrix} \right) \\
 \label{rho-red-pol}
 \equiv\left(
 \begin{matrix}
x & z\\
 z^* & 1-x
 \end{matrix}
 \right),
\end{gather}
where
\begin{equation}
 \label{x-z}
 x=|C_1|^2+\frac{|B_+|^2+|B_-|^2}{2},\,z=\frac{C_1B_+^*+B_+C_4^*}{\sqrt{2}}.
\end{equation}
 Eigenvalues of the matrix (\ref{rho-red-pol}) are equal to
\begin{gather}
 \nonumber
 \lambda_\pm^{(r) pol}=\frac{1\pm\sqrt{4|z|^2+(2x-1)^2}}{2}\\
 \label{eigen-red-pol}
 = \frac{1\pm\sqrt{\left(1-|B_-|^{\,2}\right)^2-|2C_1C_4-B_+^2|^2}}{2}.
\end{gather}
The equality
\begin{equation}
 \nonumber
 4|z|^2+(2x-1)^2=\left(1-|B_-|^{\,2}\right)^2-|2C_1C_4-B_+^2|^2
\end{equation}
is proved with a simple algebra [with the use of the normalization condition (\ref{norm})] and it shows, in particular, that at any values of the ququart's parameters
\begin{equation}
 1-|B_-|^{\,2}\geq|2C_1C_4-B_+^2|.
\end{equation}

The Schmidt parameter, determined by the eigenvalues (\ref{eigen-red-pol}) is given by
\begin{gather}
  \nonumber
  K^{pol}=\frac{1}{\left(\lambda_+^{(r) pol}\right)^2+\left(\lambda_-^{(r) pol}\right)^2}\\
 \label{K-pol}
  =\frac{2}{1+(1-|B_-|^{\,2})^2-|2C_1C_4-B_+^2|^2}\,.
\end{gather}
Entropy of the reduced mixed polarization state is given by
\begin{gather}
 \nonumber
 S(\rho^{pol}_r)=1-\sum_\pm\frac{1\pm
 \sqrt{\left(1-|B_-|^{\,2}\right)^2-|2C_1C_4-B_+^2|^2}}{2}\\
 \label{S(rho-red-pol)}
 \times\log_2\left(1\pm\sqrt{\left(1-|B_-|^{\,2}\right)^2-|2C_1C_4-B_+^2|^2}\right).
\end{gather}
\subsection{Degree of polarization and the Schmidt parameter}
The reduced density matrix (\ref{rho-red-pol}) can be used also for finding the one-photon polarization vector, or vector of Stokes parameters per photon
\begin{equation}
 \label{polarization-vector}
 {\vec\xi}=Tr\left(\rho_r^{pol}{\vec\sigma}\right)=\{2Re(z),-2Im(z), 2x-1\},
\end{equation}
where ${\vec\sigma}$ is the vector of Pauli matrices. In terms of Stokes parameters the degree of polarization is determined as $P=\left|{\vec\xi}\,\right|=\frac{1}{2}\left|{\vec S}\,\right|$, where ${\vec S}$ is the biphoton polarization vector \cite{BK}. By comparing Eqs. (\ref{K-pol}) and (\ref{polarization-vector}) we find the following general relation between the degree of polarization and the Schmidt parameter
\begin{equation}
 \label{pol-Schmidt}
 P^2+2\left[1-\left(K^{pol}\right)^{-1}\right]=1.
\end{equation}
This is the analog of the relation $P^2+C^2=1$ found in Ref \cite{Archive} for pure states of biphoton qutrits, for which $2(K^{-1}-1)=C^2$. In the case of mixed states we consider here, their concurrence $C^{pol}$ (found below) is not related anymore with the Schmidt parameter $K^{pol}$. But the relation (\ref{pol-Schmidt})  between the degree of polarization and the Schmidt parameter holds good. This shows once again that for mixed states the Schmidt parameter $K$ characterizes in some way the degree of occurring classical rather than quantum correlations, because the degree of polarization is a purely classical quantity.

Note also that the degree of polarization in mixed polarization states coincides exactly with the degree of polarization of the original two-qudit quqart as a whole, $P_{mixed}=P^{(4)}=\left|Tr\left(\rho^{(4)}{\vec\sigma}\right)\right|$. This is clear because to calculate $P^{(4)}$ we have to take the same traces of the ququart's density matrix $\rho^{(4)}$ (\ref{rho-ququart-dB}), (\ref{qqrt-dm-4x4}) which reduce it first to $\rho^{pol}$ (\ref{rho-pol-mixed-general}), then to $\rho^{pol}_r$ (\ref{rho-red-pol}), and in this way to the definition (\ref{polarization-vector}) of the vector of Stokes parameters. This procedure reduces the definition of $P^{(4)}$ to that of $P_{mixed}$.

\subsection{Full density matrix: eigenvalues, entropy, and von Neumann mutual information}
Returning to the general density matrix of a mixed polarization state  (\ref{rho-pol-mixed-general}), explicitly it takes the simplest form in the basis of states $\left\{\Psi^{pol}_{HH},\;\Psi_+^{{(Bell)}\,pol},\;\;\Psi^{pol}_{VV},\;\Psi_-^{{(Bell)}\,pol}\right\}$ defined in Eqs. (\ref{Bell-Psi}), (\ref{Bell-Phi}) and (\ref{HH-3 and VV-3}):
\begin{gather}
 \nonumber
 \rho^{pol}=\left(\begin{matrix}\rho^{(3)\,pol}& 0\\ 0&|B_-|^2\end{matrix}\right)\\
 \label{pol-4x4}
 =\left(
 \setlength{\extrarowheight}{0.1cm}
 \begin{matrix}
 |C_1|^2 & C_1B_+^* & C_1C_4^* & 0\\
 B_+C_1^* & |B_+|^2 & B_+C_4^* & 0\\
 C_4C_1^* & C_4B_+^* & |C_4|^2 & 0\\
 0 & 0 & 0 & |B_-|^2
 \end{matrix}
 \right),
 \end{gather}
where $\rho^{(3)\,pol}$ is the qutrit's coherence matrix \cite{DNK,B-Ch,BK}, \cite{Archive}. By comparing with the full ququart's density matrix of Eq. (\ref{qqrt-dm-4x4}) we see that averaging over frequency variables obliterates terms linear in $B_-$ and $B_-^*$. The expression on the right-hand side of Eq. (\ref{pol-4x4}) was obtained also in the paper \cite{Steinberg}, though not in the context of mixed polarization states averaged over the frequency distribution.

Eigenvalues of the density matrix $\rho^{pol}$ (\ref{pol-4x4}) can be easily found and, interesting enough, they appear to coincide with eigenvalues of the two-frequency mixed state considered above (\ref{eigenvalues-full}), $\lambda_{1,2}=0$, $\lambda_3=|B_-|^2$, and $\lambda_4=1-|B_-|^2$. Hence, the entropy $S(\rho^{pol})$ is given by the same equation as $S(\rho^{freq})$ (\ref{S(rho-freq)}) and in its dependence on $|B_-|$  is characterized by the same curve as in Fig. \ref{Fig1}. Again, the polarization state with invisible frequency variables is pure only in two cases, $B_-=0$ and $|B_-|=1$, and otherwise it is mixed.

A special note should be done about the case $|B_-|=1$ (and $C_1=B_+=C_4=0$), when the polarization density matrix of  Eq. (\ref{rho-pol-mixed-general}) is reduced to $\rho^{pol}|_{B_-=1}=\Psi_-^{(Bell) pol}\otimes\left(\Psi_-^{(Bell) pol}\right)^\dag$. This density matrix describes a pure polarization state with the antisymmetric wave function $\Psi^{pol}|_{B_-=1}=\Psi_-^{(Bell) pol}$. The result can seem to be in contradiction with the general requirement of the Boze-Einstein statistics according to which all biphoton wave functions are obliged to be symmetric. But, actually, there is no contradiction because in the case under consideration the polarization wave function $\Psi^{pol}|_{B_-=1}$ is only one part of a more general polarization-frequency biphoton wave function $\Psi_{BPh}|_{B_-=1}=\Psi_-^{(Bell) pol}\otimes\Psi_-^{(Bell) freq}$, which is symmetric. In a more general formulation, to produce antisymmetric pure polarization Bell state $\Psi_-^{(Bell) pol}$ one has to produce, in fact, a more general symmetric state with at least one degree of freedom additional to polarization such that ($i$) the total wave function is factorized, i.e, is given by a product of parts depending on  polarization and other degree(s) of freedom and ($ii$) both parts are antisymmetric. Then by ignoring the additional degree(s) of freedom one can observe the antisymmetric pure polarization state $\Psi_-^{(Bell) pol}$. In the case of the polarization-frequency biphoton ququarts, the role of such additional degree of freedom is played by photon frequencies.

The von Neumann mutual information of the mixed polarization state is determined by the usual formula
\begin{equation}
 \label{I(rho-pol)}
 I(\rho^{pol})=2S(\rho_r^{pol})-S(\rho^{pol})
\end{equation}
with $S(\rho_r^{pol})$ and $S(\rho^{pol})$ given by Eqs. (\ref{S(rho-red-pol)}) and (\ref{S(rho-freq)}).

\subsection{Concurrence of the mixed polarization states}

Concurrence of the mixed polarization state is determined by means of the standard Wootters procedure \cite{Wootters}. It's best starting from  for the density matrix $\rho^{pol}$ written in the form of Eqs. (\ref{pol-4x4}). Note also that the qutrit's coherence matrix in the first of these equations can be presented as $\rho^{(3)}=|\Psi^{(3)}\rangle\langle\Psi^{(3)}|$ with $|\Psi^{(3)}\rangle={\tiny\left(\begin{matrix}C_1\\ B_+\\C_4\end{matrix}\right)}$. The next step consists in constructing the spin-flipped and complex conjugated matrix ${\widetilde\rho}^{\,pol}$. Spin flipping means changing polarizations $H\rightleftharpoons V$ with multiplication of the wave functions by $i$ for  $H$- and $-i$ for $V$-polarization. This gives the following transformation rules for the ququart's coefficients $C_1\rightarrow -{\widetilde C}_4$, $B_+\rightarrow {\widetilde B}_+$, $C_4\rightarrow -{\widetilde C}_1$, $B_-\rightarrow -{\widetilde B}_-$. As a result we get  the following expression for the matrices ${\widetilde\rho}^{\,pol}$ and ${\widetilde\rho}^{(3)\,pol}$
\begin{gather}
 \nonumber
 {\widetilde\rho}^{\,pol}=\left(\begin{matrix}{\widetilde\rho}^{\,(3)}&0\\0&|B_-|^2\end{matrix}\right),\\
 \label{tilde-rho}
 {\widetilde\rho}^{\,(3)}
 =|{\widetilde\Psi}^{\,(3)}\rangle
 \langle{\widetilde\Psi}^{\,(3)}|,\,
 |{\widetilde\Psi}^{\,(3)}\rangle={\scriptsize\left(\begin{matrix}-C_4\\B_+\\-C_4\end{matrix}\right)}.
\end{gather}
With the help of these expressions, the product of matrices $\rho^{pol}$ and ${\widetilde\rho}^{\,pol}$ takes the form
\begin{equation}
 \label{rho x thilde rho}
 \rho^{pol}{\widetilde\rho}^{\,pol}=
 \left(\begin{matrix}{\rho}^{(3)}{\widetilde\rho}^{\,(3)}&0\\0&|B_-|^4\end{matrix}\right),
\end{equation}
where
\begin{gather}
 \nonumber
 \rho^{(3)}{\widetilde\rho}^{\,(3)}=|\Psi^{(3)}\rangle\langle\Psi^{(3)}
 |{\widetilde\Psi}^{\,(3)}\rangle\langle{\widetilde\Psi}^{\,(3)}|\\
 \nonumber
 =-(2C_1^*C_4^*-B_+^{*\,2})
 |\Psi^{(3)}\rangle\langle{\widetilde\Psi}^{\,(3)}|\\
 \label{rho3 x rho3}
 =-(2C_1^*C_4^*-B_+^{*\,2})\left(
 \setlength{\extrarowheight}{0.1cm}
 \begin{matrix}
 -C_1C_4&C_1B_+&-C_1^2\\
 -B_+C_4&B_+^2&-B_+C_1\\
 -C_4^2&C_4B_+&C_4C_1
 \end{matrix}
 \right).
\end{gather}
Eigenvalues of the matrix $\rho^{pol}{\widetilde\rho}^{\,pol}$ are given by $|B_-|^4$ plus three eigenvalues of the 3D matrix $\rho^{(3)}{\widetilde\rho}^{\,(3)}$. Eigenvalues of the last matrix can be found explicitly to yield
\begin{equation}
 \label{eigen-rho-rho}
 \lambda_{\rho{\widetilde\rho}}^{(1)}
 =\lambda_{\rho{\widetilde\rho}}^{(2)}=0,\;
 \lambda_{\rho{\widetilde\rho}}^{(3)}=|2C_1C_4-B_+^2|^2,\;
 \lambda_{\rho{\widetilde\rho}}^{(4)}=|B_-|^4.
\end{equation}
Square roots of these values are just the Wootters eigennumbers:
 \begin{equation}
 \label{eigen-W}
 \lambda_{W}^{(1)}
 =\lambda_{W}^{(2)}=0,\;
 \lambda_{W}^{(3)}=|2C_1C_4-B_+^2|,
 \lambda_{W}^{(4)}=|B_-|^2.
\end{equation}
Concurrence of the mixed polarization state is given by the difference between  $\left[\lambda_W^{(i)}\right]_{\max}$
and the sum of all other $\lambda_W^{(i)}$
\begin{equation}
 \label{concurrence-pol}
 C^{pol}=\left||2C_1C_4-B_+^2|-|B_-|^2\right|.
\end{equation}

\subsection{Example 1}

 $C_1=C_4=0$. This is the case when the original ququart's wave function has the form of a sum of two double-Bell states
 \begin{gather}
  \nonumber
  \Psi^{(4)}=B_+\Psi_+^{(Bell)\,pol}\otimes\Psi_+^{(Bell)\,freq}\\
  \label{Two-doubleBells}
  +B_-\Psi_+^{(Bell)\,pol}\otimes\Psi_-^{(Bell)\,freq}
 \end{gather}
 with $|B_+|^2+|B_-|^2=1$. The mixed-state polarization density matrix $\rho^{pol}$ (\ref{rho-pol-mixed-general}) corresponding to this wave function is reduced in this case to the form identical to that of the mixed-state frequency density matrix $\rho^{freq}$ (\ref{freq-mixed-rho}):
\begin{gather}
 \nonumber
 \rho^{pol}=(1-|B_-|^2)\Psi_+^{(Bell)\, pol}\otimes\left(\Psi_+^{(Bell)\, pol}\right)^\dag\\
 \label{rho-C1=C2=0}
 +|B_-|^2\Psi_-^{(Bell)\, pol}\otimes\left(\Psi_-^{(Bell)\, pol}\right)^\dag.
\end{gather}
Consequently, the states (\ref{rho-C1=C2=0}) are characterized by the same correlation parameter as shown in Fig. \ref{Fig2}. In particular, in this case $K^{pol}\equiv 2$ and $C^{pol}=\left|1-2|B_-|^2\right|$. Besides, owing to the relation between the degree of polarization and the Schmidt parameter (\ref{pol-Schmidt}), $P\equiv 0$, i.e. in this case all states (\ref{rho-C1=C2=0}) are unpolarized, independently of values of the complex constants  $B_+$ and $B_-$.

\subsection{Example 2}
 $B_+=0$. In this case the density matrix $\rho^{pol}$ (\ref{rho-pol-mixed-general}) takes the form
\begin{gather}
 \nonumber
 \rho^{pol}=\big[C_1\Psi_{HH}^{pol}+C_4\Psi_{VV}^{pol}\big]\otimes\big[C_1\Psi_{HH}^{pol}+C_4\Psi_{VV}^{pol}\big]^\dag\\
 \label{rho-B+=0}
 + |B_-|^2\Psi_-^{(Bell)\,pol}\otimes\left(\Psi_-^{(Bell)\,pol}\right)^\dag,
\end{gather}
where $\Psi_{HH}^{pol}$ and $\Psi_{VV}^{pol}$ are given by Eqs. (\ref{HH-3 and VV-3}).
Expressions for the nonzero eigenvalues of the reduced density matrix $\lambda_\pm^{(r)pol}$, Schmidt parameter $K^{pol}$, degree of polarization $P$, and concurrence $C^{pol}$ follow from general Eqs. (\ref{eigen-red-pol}), (\ref{K-pol}), (\ref{pol-Schmidt}), (\ref{concurrence-pol})
\begin{gather}
 \nonumber
 \lambda_\pm^{(r)pol}=\frac{1\pm\big||C_1|^2-|C_4|^2\big|}{2},\,
 K^{pol}=\frac{2}{1+\left(|C_1|^2-|C_4|^2\right)^2},\\
 \label{B+=0}
 P=\big||C_1|^2-|C_4|^2\big|,\,C^{pol}=\big|\left(|C_1|+|C_4|\right)^2-1\big|
\end{gather}
with the normalization condition $|C_1|^2+|C_4|^2=1-|B_-|^2$. These expressions are two-parametric, and for their simpler analysis let us consider two one-parametric cases.

\subsubsection{$B_+=C_4=0, \;|C_1|^2=1-|B_-|^2$}
This case corresponds to the following original ququart's wave function (\ref{ququart-dB})
\begin{gather}
 \Psi^{(4)}=C_1\Psi_{HH}^{(4)}
 \label{wf-B+=C4=0}
 + B_-\Psi_-^{(Bell)\,pol}\otimes\Psi_-^{(Bell)\,freq},
\end{gather}
or in the traditional form (\ref{ququart}) to coefficients $C_2=B_-/\sqrt{2}$ and $C_3=-B_-/\sqrt{2}$:
\begin{gather}
 %\nonumber
 \Psi^{(4)}=C_1\Psi_{HH}^{(4)}
 \label{wf-B+=C4=0-tradit}
 +\frac{ B_-}{\sqrt{2}}\left(\Psi_{HV}^{(4)}-\Psi_{VH}^{(4)}\right).
\end{gather}

Under these conditions the density matrix of the mixed polarization state (\ref{rho-pol-mixed-general}), (\ref{rho-B+=0}) takes the form
\begin{gather}
 \nonumber
 \rho^{pol}=(1-|B_-|^2)\Psi_{HH}^{pol}\otimes\left(\Psi_{HH}^{pol}\right)^\dag\\
 + |B_-|^2\Psi_-^{(Bell)\,pol}\otimes\left(\Psi_-^{(Bell)\,pol}\right)^\dag ,
  \label{rho-B+=C4=0}
\end{gather}
and Eqs. (\ref{B+=0}) are reduced to
\begin{gather}
 \nonumber
 \lambda_\pm^{(r)pol}=\frac{1\pm\big|1-|B_-|^2\big|}{2},\;
 K^{pol}=\frac{2}{2-2|B_-|^2+|B_-|^4}\;,\\
 \label{B+=C4=0}
 P=1-|B_-|^2=1-C^{pol},\;C^{pol}=|B_-|^2.
\end{gather}
Note that the derived relation between the degree of polarization and concurrence $P=1-C^{pol}$ is specific for the case under consideration, $B_+=C_4=0$. But this result is rather important because it shows that the degree of entanglement can be found in this case via direct experimental measurements of the degree of polarization of  the state (\ref{rho-B+=C4=0}).

\begin{figure}[h]
\centering\includegraphics[width=7cm]{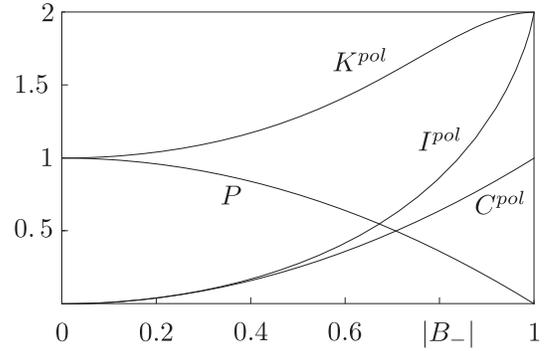}
\caption{{\protect\footnotesize {The Schmidt parameter $K^{pol}$, concurrence $C^{pol}$, von Neumann mutual information $I^{pol}$ and the degree of polarization $P$ as functions of the mixing parameter $B_-$ for the state   (\ref{rho-B+=C4=0}) }.}}\label{Fig3}
\end{figure}

At $B_+=C_4=0$ the entropy of the reduced state (\ref{rho-B+=C4=0}) equals to
\begin{gather}
 \nonumber
 S(\rho^{pol}_r)=-\sum_\pm\lambda_\pm^{(r)pol}\log_2(\lambda_\pm^{(r)pol})\\
 \label{entr-pol-b1}
 =1-\sum_\pm\frac{1\pm\big|1-|B_-|^2\big|}{2}\log_2(1\pm\big|1-|B_-|^2\big|).
\end{gather}
The functions $K^{pol}(|B_-|)$, $C^{pol}(|B_-|)$ and $P(|B_-|)$ are shown in Fig. \ref{Fig3} together with the von Neumann mutual information $I(|B_-|)=2S(\rho^{pol}_r)-S(\rho^{pol})$ determined by Eqs. (\ref{S(rho-freq)}) and (\ref{entr-pol-b1}). The left and right borders of this Figure correspond to pure two-qubit polarization states,
$\Psi^{pol}_{HH}$ at $B_- =0$ and $\Psi_-^{(Bell)\,pol}$ at $|B_-|=1$. The state $\Psi^{pol}_{HH}$ is maximally polarized and has no correlations at all, either quantum or classical ones. The state $\Psi_-^{(Bell)\,pol}$ is unpolarized and has maximal possible degrees of classical and quantum correlations, $C=1$ and $C_{cl}=I-C=1$. At intermediate values of $|B_-|$, $1>|B_-|>0$ the polarization state (\ref{wf-B+=C4=0}), (\ref{wf-B+=C4=0}) is mixed and possesses both quantum and classical correlations. At $|B_-|\neq 0, 1$ the degree of classical correlations $C_{cl}$ is somewhat larger than  $I(|B_-|)-C(B_I)$ because the curve of the relative entropy $S_{rel}$ (not shown here) inevitably is located somewhat lower than $C(|B_-|)$.

\subsubsection{$B_+=0$, $|C_1|=|C_4|=\sqrt{\frac{1-|B_-|^2}{2}}$}
In this case the general density matrix of the mixed polarization state (\ref{rho-pol-mixed-general}) has the form
\begin{gather}
 \nonumber
 \rho^{pol}=\frac{1-|B_-|^2}{2}\left[e^{i\varphi_1}\Psi_{HH}^{pol}+e^{i\varphi_4}\Psi_{VV}^{pol}\right]\\
 \nonumber
 \otimes\left[e^{i\varphi_1}\Psi_{HH}^{pol}+e^{i\varphi_4}\Psi_{VV}^{pol}\right]^\dag\\
 \label{wf-c1=c4}
 +|B_-|^2\Psi_-^{Bell,\,pol}\otimes\left(\Psi_-^{Bell,\,pol}\right)^\dag,
\end{gather}
where $\varphi_{1,4}$ are phases of the coefficients $C_{1,4}$. In terms of entanglement and correlations, the state (\ref{wf-c1=c4}) is equivalent to the two-frequency mixed state (\ref{freq-mixed-rho}) and to the above considered case $C_1=C_4=0,\, B_+\neq 0$ (\ref{rho-C1=C2=0}). Eqs. (\ref{B+=0}) are reduced to
\begin{gather}
  \lambda_\pm^{(r)pol}=\frac{1}{2},\; K^{pol}=2,
 \label{B+=0 C1=C4}\;P=0,\;C^{pol}=\big|1-2|B_-|^2\big|,
\end{gather}
and all correlation parameters as functions of $|B_-|$ are characterized by the same curves as in Fig. \ref{Fig2}.

\section{Comparison with the two-qubit model of biphoton ququarts}
The consideration given above assumes experimental investigation of ququarts with devices nonselective either with respect to frequencies or polarizations of photons. In the case of spectral mixed states the detectors and beam splitters to be used are assumed to be spectrally non-selective. On the other hand, there is an alternative method of measurements. It's possible to split the biphoton beam for two channels by a dichroic beam splitter to get two beams with only high-frequency or only low-frequency photons. After this it's possible to investigate spectral correlations only in one of these two channels independently of the other one.

A theoretical picture corresponding to this procedure is based on the observation that the general wave function of a biphoton ququart (\ref{ququart}) can be written in the ``coordinate$"$ representation in the form
\begin{gather}
 \nonumber
 \Psi^{(4)}(\sigma_1,\omega_1;\sigma_2,\omega_2)=\\
 \label{wf-2-terms}
 =\Psi^{(h)}_{2\,qb}(\sigma_1,\sigma_2)
 \delta_{\omega_1,\omega_h}\delta_{\omega_2,\omega_l}
 +\Psi_{2\,qb}^{(l)}(\sigma_1,\sigma_2)
 \delta_{\omega_1,\omega_l}\delta_{\omega_2,\omega_h},
\end{gather}
where $\Psi^{(h)}_{2\,qb}$ and $\Psi^{(l)}_{2\,qb}$ are the often used two-qubit purely polarization ququart's wave functions with a broken symmetry
\begin{gather}
 \nonumber
 \Psi^{(h)}_{2\,qb}(\sigma_1,\sigma_2)=C_1\delta_{\sigma_1,H}\delta_{\sigma_2,H}
 +C_2\delta_{\sigma_1,H}\delta_{\sigma_2,V}\\
 \nonumber
 +C_3\delta_{\sigma_1,V}\delta_{\sigma_2,H}+C_4\delta_{\sigma_1,V}\delta_{\sigma_2,V}\\
 \nonumber
 =C_1\left(\begin{matrix}1\\0\end{matrix}\right)_1^{pol}\otimes\left(\begin{matrix}1\\ 0\end{matrix}\right)_2^{pol}+C_2\left(\begin{matrix}1\\0\end{matrix}\right)_1^{pol}\otimes
 \left(\begin{matrix}0\\1\end{matrix}\right)_2^{pol}\\
 \label{psi-2-qb}
 +C_3\left(\begin{matrix}0\\1\end{matrix}\right)_1^{pol}\otimes
 \left(\begin{matrix}1\\0\end{matrix}\right)_2^{pol}+
 C_4\left(\begin{matrix}0\\1\end{matrix}\right)_1^{pol}
 \otimes\left(\begin{matrix}0\\1\end{matrix}\right)_2^{pol}.
\end{gather}
and the same for $\Psi^{(l)}_{2\,qb}$ with the replacement $C_2\rightleftharpoons C_3$.

Such wave functions, asymmetric with respect to the variable transposition $1\rightleftharpoons 2$, cannot exist by themselves, without multiplication by frequency parts in the total wave function (\ref{wf-2-terms}). But it's possible to think that the use of a dichroic beam splitter in experiments can give sense even for such ``non-existing$"$ wave functions. Indeed, in the density matrix $\rho_r$ arising from the representation (\ref{wf-2-terms}) the off-diagonal terms disappear and $\rho_r$ takes the form
\begin{gather}
 \nonumber
 \rho_r(\sigma_1,\omega_1;\sigma_1^\prime,\omega_1^\prime)=Tr_{\sigma_2,\omega_2}\rho^{(4)}=\\
 \label{red-high-low}
 \rho_{r,h}(\sigma_1,\sigma_1^\prime)\delta_{\omega_1,\omega_h}\delta_{\omega_1^\prime,\omega_h}+
 \rho_{r,l}(\sigma_1,\sigma_1^\prime)\delta_{\omega_1,\omega_l}\delta_{\omega_1^\prime,\omega_l},
\end{gather}
where
\begin{gather}
 \nonumber
  \rho_{r,\,h}=\sum_{\sigma_2}\Psi^{(h)}_{2\,qb}(\sigma_1,\sigma_2)\Psi^{(h)*}_{2\,qb}(\sigma_1^\prime,\sigma_2)\\
  \label{rho-h}
  =\left(
  \setlength{\extrarowheight}{0.1cm}
  \begin{matrix}
  |C_1|^2+|C_2|^2&C_1C_3^*+C_2C_4^*\\C_1^*C_3+C_2^*C_4&|C_3|^2+|C_4|^2
  \end{matrix}\right)
\end{gather}
and
\begin{gather}
 \nonumber
 \rho_{r,\,l}=\sum_{\sigma_2}\Psi^{(l)}_{2\,qb}(\sigma_1,\sigma_2)\Psi^{(l)*}_{2\,qb}(\sigma_1^\prime,\sigma_2)\\
  \label{rho-l}
  =
  \left(
  \setlength{\extrarowheight}{0.1cm}
  \begin{matrix}
  |C_1|^2+|C_3|^2&C_1C_2^*+C_3C_4^*\\
  C_1^*C_2+C_3^*C_4&|C_2|^2+|C_4|^2
  \end{matrix}
  \right).
\end{gather}
A half sum of these partial reduced density matrices give exactly the above derived reduced density matrix of the mixed polarization state (\ref{rho-red-pol})
\begin{equation}
 \label{half-sum}
 \rho^{pol}_r=\frac{\rho_{r,\,h}+\rho_{r,\,l}}{2}.
\end{equation}

Separation of high- and low-frequency parts in the reduced density matrix $\rho_r^{(4)}$
(\ref{red-high-low}) gives rise to a possibility  of interpreting each of two terms in this equation as describing correlations in each of two channels after the dichroic beam splitter, independently of another. Note however that separation for low and high frequencies arises only in the reduced but not in the full density matrix. The latter describes all biphoton beam as a whole, either  split in a dichroic beam splitter or not.

Eigenvalues of the density matrices $\rho_{r,h}$ and $\rho_{r,l}$ (\ref{rho-h}), (\ref{rho-l}) are easily found to give the well known concurrence and Schmidt parameter of a model two-qubit ququart \cite{Bogd}
\begin{gather}
 \label{conc-2-qb}
 C_{2\,qb}=2|C_1C_4-C_2C_3|=\left|2C_1C_4-B_+^2+B_-^2\right|,\\
 \label{K-2-qb}
 K_{2\,qb}=\frac{2}{2-C_{2\,qb}^2}=\frac{2}{2-\left|2C_1C_4-B_+^2+B_-^2\right|^2},
\end{gather}
whereas in mixed polarization states these parameters are given by Eqs. (\ref{concurrence-pol}) and (\ref{K-pol}):
\begin{equation}
 \label{C-pol-again}
 C^{pol}=\left||2C_1C_4-B_+^2|-|B_-|^2\right|,
\end{equation}
\begin{gather}
  \label{K-pol-again}
  K^{pol}=\frac{2}{1+(1-|B_-|^{\,2})^2-|2C_1C_4-B_+^2|^2}\,.
\end{gather}

Differences between Eqs. (\ref{conc-2-qb}), (\ref{K-2-qb}) and (\ref{C-pol-again}), (\ref{K-pol-again}) reflect a general difference between the mixed polarization state of the ququart as a whole and its part in separate high- or low-frequency channel arising in the reduced density matrix. Probably, the simplest way of seeing these differences in experiments consists in measuring the degree of polarization in the schemes with and without the dichroic beam splitter. In both cases the degree of polarization is related to the corresponding Schmidt parameter:
\begin{gather}
 \label{pol-Schmidt-2-qb-mixed}
 P^{(4)\,2}+2\left[1-\frac{1}{K^{pol}}\right]=1,\,
 P_{2\,qb}^2+2\left[1-\frac{1}{K_{2\,qb}}\right]=1.
\end{gather}
Specifically, in two simple cases of subsections {\bf E} and {\bf F}, section {\bf V}, Eqs. (\ref{conc-2-qb})-(\ref{pol-Schmidt-2-qb-mixed}) give the following results.

\begin{figure}[h]
\centering\includegraphics[width=7cm]{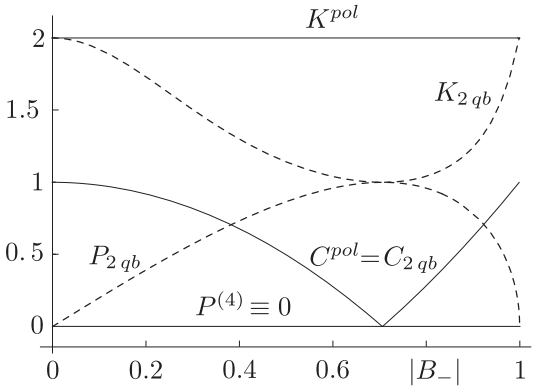}
\caption{{\protect\footnotesize {The Schmidt parameter $K^{pol}$ of the mixed state (\ref{rho-C1=C2=0}) and the degree of polarization $P^{(4)}$ of the state $\Psi^{(4)}$ (\ref{wf-C1-C4-0}) (solid lines) and the same quantities for the two-qubit state (\ref{psi-2-qb}) (dashed lines) with coinciding curves for concurrencies $C^{pol}$ and $C_{2\,qb}$ (solid line) for the case $C_1=C_4=0$.}}}\label{Fig4}
\end{figure}

1) $C_1=C_4=0$ with $|B_+|^2+|B_-|^2=1$, which corresponds to the original ququart's wave function
\begin{equation}
 \label{wf-C1-C4-0}
 \Psi^{(4)}=\frac{B_++B_-}{\sqrt{2}}\Psi^{(4)}_{HV}+\frac{B_+-B_-}{\sqrt{2}}\Psi^{(4)}_{VH}.
\end{equation}
In this case $C_{2\,qb}=|B_+^2-B_-^2|$, $P_{2\,qb}=\sqrt{1-|B_+^2-B_-^2|^2}$ whereas $K^{pol}\equiv2$, $C^{pol}=\left||B_+|^2-|B_-|^2\right|$, and $P^{(4)}\equiv 0$, and the dependence of all these parameters on $|B_-|$ is shown in Fig. \ref{Fig4} for real values of $B_+$ and $B_-$.
 The differences between the curves corresponding to the ququart's mixed state and to the two qubit state are rather well pronounced in the case of the Schmidt parameter and the degree of polarization whereas the curves for the concurrences $C^{pol}$ and $C_{2\,qb}$ coincide. This last coincidence is accidental and, under the same conditions $C_1=C_4=0$, occurs only owing to the assumption that both constants $B_+$ and $B_-$ are real. If, for example,  the constant $B_+$ is complex rather than real, $B_+=e^{i\varphi}|B_+|$ (with real $B_-$), then the concurrence of the mixed polarization state (\ref{C-pol-again}) remains the same as shown in Fig. (\ref{Fig4}), $C^{pol}=\left|1-2|B_-|^2\right|$, whereas the two-qubit concurrence (\ref{conc-2-qb}) becomes equal to $C_{2\,qb}=\left[(1-|B_-|^2)^2-2(1-|B_-|^2)|B_-|^2\cos 2\varphi+|B_-|^4\right]^{1/2}$, and $C^{pol}\neq C_{2\,qb}$ if only $\varphi\neq 0\;{\rm or}\;\pi$.

\begin{figure}[h]
\centering\includegraphics[width=7cm]{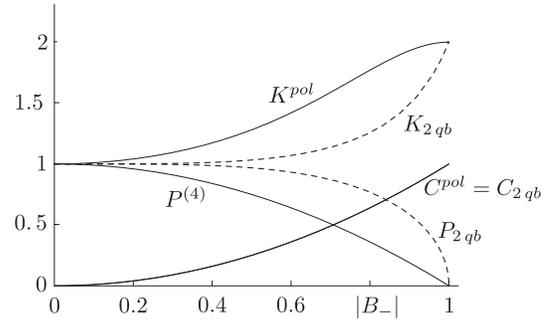}
\caption{{\protect\footnotesize {The same as in Fig. \ref{Fig4} but for $C_4=B_+=0$.}}}\label{Fig5}
\end{figure}

2) $C_4=B_+=0$ with $|C_1|^2+|B_-|^2=1$. The corresponding original ququart's wave function is given by Eq. (\ref{B+=C4=0}). The Schmidt parameter, concurrence and degree of polarization found in the two-qubit model are equal now to ${\displaystyle K_{2\,qb}=\frac{2}{2-|B_-|^4}}$, $C_{2\,qb}=|B_-^2|$, and $P_{2\,qb}=\sqrt{1-|B_-|^4}$,  whereas in the case of mixed states we get ${\displaystyle K^{pol}=\frac{2}{2-2|B_-|^{\,2}+|B_-|^{\,4}}}$, $C^{pol}=|B_-|^2$, $P^{(4)}=1-|B_-|^2$. In this case, again, the concurrences $C^{pol}$ and $C_{2\,qb}$ coincide with each other and again because of a specific choice of the ququart's constants $C_4=B_+=0$. But the degree of polarization $P^{(4)}$ does not coincide with $P_{2\,qb}$ and the Schmidt parameter $K^{pol}$ does not coincide with $K_{2\,qb}$. Dependence of all these quantities on the mixing parameter $|B_-|$ is shown in Fig. \ref{Fig5}.
The difference between the curves for the Schmidt parameter $K(|B_-|)$ and the degree of polarization $P(|B_-|)$ in cases of mixed state and of the two-qubit model is rather well pronounced and experimentally measurable.

\section{Conclusion}

In Conclusion, biphoton polarization-frequency ququarts are pure states of two photons with two degrees of freedom: polarization and frequency. Owing to indistinguishability of photons, their total wave functions
must be symmetric with respect to the particle transpositions. This gives rise to the intrinsic entanglement of such states related to their symmetry (symmetry entanglement). In addition, entanglement can arise also if the ququarts are formed by a superposition of several occupation-number configurations (configuration entanglement). Both symmetry and configuration entanglement are summed in the total entanglement of ququarts characterized by their Schmidt decomposition and Schmidt parameter $K$. If, however, either frequency- or polarization-variables are considered as ``invisible$"$, the full density matrix of ququarts has to be averaged over these invisible variables. As a result one gets mixed two-qubit states depending only on either polarization or frequency variables. Such mixed states have both classical and quantum correlations, and parameters characterizing levels of these correlations are found and analyzed. In the case of mixed states the Schmidt parameter $K$ is shown to be not related to the degree of entanglement. But it remains useful as characterizing the degree of polarization of ququarts. The degree of polarization of ququarts is shown to be identical to the degree of polarization occurring in mixed polarization states arising after averaging of the full quqart's density matrix over the frequency variables. This degree of polarization can be found experimentally by means of a direct measurement of the Stokes parameters of the original ququart biphoton beam not split for two parts by DBS. On the other hand, in experiments with DBS, by measuring parameters of only one part of a split ququart's beam, one can find the degree of polarization of such half a beam, different from the degree of polarization of the whole ququart's beam. Such measurement permit to see two-qubit features of ququarts but they do not determine the degree of polarization of the ququart beam as a whole. In two simple cases the difference between measurements to be done with and without using the DBS is illustrated by pictures of Figs. \ref{Fig4} and \ref{Fig5} which show that the difference is rather well pronounced and is accessible for experimental control.

\begin{acknowledgments}
Authors are grateful to S.P. Kulik and S.S. Straupe for multiple fruitful discussions. The work is supported partially by the grants of the Russian Foundation for Basic Research 02-08-10404 and 02-11-01043.
\end{acknowledgments}

\end{document}